\documentclass[aps,prb,reprint,floatfix,citeautoscript,longbibliography,superscriptaddress]{revtex4-1}
\usepackage{graphicx}
\usepackage{bm,amsmath,amssymb,mathrsfs,dcolumn}
\usepackage[usenames]{color}
\usepackage[colorlinks=true, linkcolor=blue, citecolor=blue, urlcolor=blue, linktoc=page, bookmarks=false, pdfstartview={FitH}, pdfborder={0 0 0.0 [3 3]}]{hyperref}
\hypersetup{pdfborder=0 0 0,colorlinks=true,citecolor=blue,linkcolor=blue}
\usepackage{subfigure}
\usepackage{ulem}

\begin{document}
\title{Proper dissipative torques in antiferromagnetic dynamics}
\author{H. Y. Yuan}
\affiliation{Department of Physics, Southern University of Science
and Technology of China, Shenzhen, Guangdong, 518055, China}
\author{Qian Liu}
\affiliation{The Center for Advanced Quantum Studies and Department
of Physics, Beijing Normal University, Beijing 100875, China}
\author{Ke Xia}
\affiliation{The Center for Advanced Quantum Studies and Department
of Physics, Beijing Normal University, Beijing 100875, China}
\affiliation{Synergetic Innovation Center for Quantum Effects and
Applications (SICQEA), Hunan Normal University, Changsha 410081, China}
\author{Zhe Yuan}
\email[Corresponding author: ]{zyuan@bnu.edu.cn}
\affiliation{The Center for Advanced Quantum Studies and Department
of Physics, Beijing Normal University, Beijing 100875, China}
\author{X. R. Wang}
\email[Corresponding author: ]{phxwan@ust.hk}
\affiliation{Department of Physics, The Hong Kong University of
Science and Technology, Clear Water Bay, Kowloon, Hong Kong}
\affiliation{HKUST Shenzhen Research Institute, Shenzhen 518057, China}
\date{\today}

\begin{abstract}
There is little doubt that the magnetization dynamics of ferromagnetic systems
is governed by the Landau-Lifshitz-Gilbert equation or its generalization with
various spin torques. In contrast, there are several sets of dynamic equations
for two-sublattice antiferromagnets (AFMs) in literature that have different
forms of dissipative torques and no proper dynamic equations for
multi-sublattice AFMs and ferrimagnets in general. Here we introduce the
general Rayleigh dissipation functional into the Lagrange equation and derive
the proper form of the dissipative torques in the phenomenological equations
for the AFMs with multiple sublattices. A new type of dissipative torque
arising from inter-sublattice drag effect is discovered that has important
influences on magnon lifetime and domain wall motion. In particular,
our theory unifies different dynamic equations of AFMs in literature.
\end{abstract}

\maketitle
\section{Introduction}
There is a reviving interest in antiferromagnetic physics \cite
{MacDonald, ke2008,Jungwirth,Hals2011,Tveten2014,Cheng,Kim2014,
Troncoso2015,Shiino2016,Helen2016,yuan2017} since the discoveries
of spin-transfer torque \cite{MacDonald,ke2008} and anisotropic
magnetoresistance \cite{Jungwirth} in antiferromagnets (AFMs).
These discoveries make AFMs promising spintronics materials for
data storage and information processing besides their traditional
usage as pinning materials because AFMs have no stray field
and their resonance frequency are in terahertz (THz) range
\cite{Kampfrath2010} so that AFM devices have no cross-talking
problem and can operate at high speed \cite{Jungwirth}.
The future development and application of AFM devices rely on our
comprehensive understanding of AFM dynamics, which are fundamentally
different from ferromagnetic dynamics not only at the quantum
mechanical level, but also at the classical physics level.
At the quantum level, it is impossible to use a unitary
transformation to map them from one to the other.
At the classical level, a ferromagnet (FM) can be described by
magnetization $\bf {m}$ while an AFM should be described by at
least two order parameters, e.g. the magnetization of each sublattice
$\mathbf{m}_1$ and $\mathbf{m}_2$ that are often redefined as
the N\'{e}el order $\mathbf{n}\equiv \mathbf{m}_1-\mathbf{m}_2$ and
the net magnetization $\mathbf{m} \equiv \mathbf{m}_1+ \mathbf{m}_2$.
Thus one should not be surprised if their dynamics are different.

The magnetization dynamics of a FM is governed by the
Landau-Lifshitz-Gilbert (LLG) equation \cite{Landau1935,Gilbert2004},
\begin{equation}
\dot{\mathbf{m}}= -\mathbf{m} \times  \mathbf{h}
+ \alpha \mathbf{m} \times \dot{\mathbf{m}},
\label{llg}
\end{equation}
where $\mathbf{h}$ is the effective field consisting of exchange field,
anisotropy field and external field and $\alpha$ is damping constant.
The first term on the right hand side of Eq. (\ref{llg}) describes
the precessional motion of magnetization around its effective field.
The second term is the Gilbert damping that forces the
magnetization to align with the effective field \cite{Gilbert2004}.
The correctness of the LLG equation for the magnetization dynamics
of FMs was verified by the good agreement between experiments
\cite{experiments} and theories \cite{theory}, and there is little
doubt about general applicability of the LLG equation for FMs.

Things are very different for AFMs. Despite of many attempts,
there is no consensus on the dynamic equations of AFMs,
where the proper form of the dissipation is particularly unclear.
So far, there are different sets of equations used in studying
magnetization dynamics of two-sublattice AFMs.
In 1950s, Kittel and coworkers introduced the coupled
Landau-Lifshitz equations on each sublattice to describe
the antiferromagnetic resonance \cite{Kittel1951, Keffer1952}.
This set of equations (with Gilbert damping) has also been used
recently to study spin-transfer torque, spin wave excitation and domain
wall (DW) dynamics \cite{Helen2014,ke2008,Kampfrath2010,Selzer2016}.
Later, Bar'yakhtar {\it et al.} proposed a phenomenological theory to
include both the longitudinal and transverse relaxation of magnetic
moments based on the assumption of the magnetization conservation
\cite{Baryakhtar1984,Baryakhtar1988,Baryakhtar201301,Baryakhtar201302}.
This assumption does not allow the relative motion of two sublattices,
resulting in zero damping for the motion of the N{\'e}el order.
This result is not supported by other theories in literature
\cite{Helen2014, Hals2011, Tveten2016,Atxitia2017}. In addition, recent
first-principles calculation has definitely proved non-zero damping
associated with the motion of the N{\'e}el order \cite{ly2017}.
The Baryakhtar's approach was further pursued by Gomonay and
coworkers \cite{Gomonay2012} to construct the dissipation function
in antiferromagnetic dynamics. They also treated the spin vectors
of an AFM as a rigid-body rotation without the relative motion
of sublattices in the same manner as Baryakhtar {\it et al}.
More recently, an alternative set of equations is derived from
the Lagrange equation of an AFM, in which the dissipative torques
are phenomenologically introduced \cite{Hals2011, Tveten2016}.
Specifically, $\alpha_m$ and $\alpha_n$ are defined as the damping
coefficients for the motion of magnetization $\mathbf{m}$ and the
N\'{e}el order $\mathbf{n}$, respectively.
The resulting equations of this approach are later used to investigate
the AFM dynamics by assuming $\alpha_m =\alpha_n$ \cite{Tveten2013,
Troncoso2015,Cheng,Shiino2016} or $\alpha_m=0$ \cite{Takei2014}.
As it was pointed out in recent review articles \cite{Helen2014,
Atxitia2017}, determining the quantitative values of the damping
coefficients and their physical mechanisms arising from the exchange
interaction or the relativistic origin remains a challenge in the field
of magnetism. Besides the open questions of the proper dissipative
torques for two-sublattice AFMs, no convincing dynamic equations exist in
literature for the AFMs with three or more sublattices or ferrimagnets.
For the latter case, the net magnetization at equilibrium is no zero.
The above unsolved issues motivate the current work.

In this letter, we consider a general AFM with $N$ sublattices that may
not be collinear with each other. By introducing the proper Rayleigh
dissipation functional into the Lagrange equation, we derive new
dissipative torques resulting from the inter-sublattice drag effect.
This new torque has the anti-damping characteristic and increases the
magnon lifetime in AFMs. Releasing the (improper) constraints used in
literature, our new AFM dynamic equations essentially unify all the
previously equations, which are different with one another.
In addition, our results can naturally explain the recent
first-principles calculation of the damping parameters in AFMs.

\section{General theory}
We consider an $N$-sublattice AFM with sublattice
magnetization $\mathbf{m}_1,\mathbf{m}_2,\cdots,\mathbf{m}_N$.
The Lagrangian functional of the AFM depends on $\mathbf m_i$ and their
time derivative $\dot{\mathbf{m}}_i \equiv\partial_t \mathbf{m}_i$, i.e.
$\mathcal{L}=\mathcal{L}(\mathbf{m}_1,\mathbf{m}_2,\cdots,\mathbf{m}_N,
\dot{\mathbf{m}}_1,\dot{\mathbf{m}}_2,\cdots,\dot{\mathbf{m}}_N)$.
The dissipation is described by the Rayleigh dissipation functional
$\mathcal{R}=\mathcal{R}(\dot{\mathbf{m}}_1,
\dot{\mathbf{m}}_2,\cdots,\dot{\mathbf{m}}_N)$ \cite{Goldstein,note_nu}.
Then the Lagrange equation with the dissipation term is given by
\begin{equation}
-\frac{\delta \mathcal{L}}{\delta \mathbf{m}_i}+ \frac{\partial }
{\partial t} \frac{\delta \mathcal{L}}{\delta \dot{\mathbf{m}}_i}
+\frac{\delta \mathcal{R}}{\delta \dot{\mathbf{m}}_i} =0.
\label{lagrangian}
\end{equation}
The Lagrangian, $\mathcal{L}=\mathcal{T}-\mathcal{U}$, consists
of the kinetic energy density functional $\mathcal{T}$
and the potential energy density functional $\mathcal{U}$.
Thus the Lagrange Eq.~\eqref{lagrangian} can be recast as
\begin{equation}
-\frac{\delta \mathcal{T}}{\delta \mathbf{m}_i}+ \frac{\partial }
{\partial t}\frac{\delta \mathcal{T}}{\delta \dot{\mathbf{m}}_i}+
\left(-\mathbf{h}_i+\frac{\delta\mathcal{R}}{\delta\dot{\mathbf{m}}_i}
\right )=0,
\label{lag}
\end{equation}
where $\mathbf{h}_i =-\delta \mathcal{U}/\delta \mathbf{m}_i$ is the
effective magnetic field acting on the $i$-th sublattice.
We have used the fact that $\mathcal U$ depends only on the magnetization
$\mathbf m_i$ and hence $\delta\mathcal U/\delta\dot{\mathbf m}_i=0$.

The kinetic energy of a spin comes from the Berry phase caused
by spin motion \cite{Tatara2008}, i.e. $\mathcal{T} \equiv i \langle
\mathbf{m}(t)| \partial_t |\mathbf{m}(t) \rangle$.
For a FM, the kinetic energy can be rewritten in a coordinate invariant
form of $\mathcal{T}=\mathbf{A}(\mathbf{m}) \cdot  \dot{\mathbf{m}}$,
where the magnetic potential $\mathbf{A}$ is determined by
$\nabla \times \mathbf{A}(\mathbf{m}) = \mathbf{m}$ \cite{notesm}.
As a natural extension to the $N$-sublattice AFM, $\mathcal{T}
=\sum_{i=1}^N \mathbf{A}(\mathbf{m}_i)\cdot  \dot{\mathbf{m}}_i$.
This is because the Berry phase induced by the variation of the
magnetization is additive for multiple sublattices $\mathbf m_i$ (See Appendix A for details).
Substituting the kinetic energy term into Eq. (\ref{lag}),
we obtain (See Appendix B for details)
\begin{eqnarray}
\dot{\mathbf{m}}_i= -\mathbf{m}_i \times \left ( \mathbf{h}_i
- \frac{\delta \mathcal{R}}{\delta \dot{\mathbf{m}}_i} \right ).
\label{llgR}
\end{eqnarray}
Here the dissipation term is essentially a ``damping field"
$-\delta \mathcal{R}/ \delta \dot{\mathbf{m}}_i$ in addition to the
effective magnetic field $\mathbf{h}_i$. The Rayleigh dissipation
functional $\mathcal{R}$ is a quadratic functional of the dynamic
variables $\dot{\mathbf{m}}_1, \dot{\mathbf{m}}_2,\cdots,\dot{
\mathbf{m}}_N$ \cite{Goldstein,Gilbert2004}, i.e.
$\mathcal{R}=(\mathbf{v}\cdot \mathbf{R} \cdot \mathbf{v}^T)/2$,
where $\mathbf{v}=( \dot{\mathbf{m}}_1, \dot{\mathbf{m}}_2,\cdots,\dot{\mathbf{m}}_N)$
and $\mathbf{R}$ is the so-called dissipation matrix.

Thus Eq.~(\ref{llgR}) in terms of $\mathbf{R}$ becomes
\begin{equation}
\dot{\mathbf{m}}_i= -\mathbf{m}_i \times\mathbf{h}_i+ \mathbf{m}_i
\times \left ( \sum_{j=1}^{N} R_{ij}\dot{\mathbf{m}}_j\right ).
\label{nsub}
\end{equation}
Equation~\eqref{nsub} governs the AFM dynamics.

Following the standard Lagrange mechanics, the energy dissipation rate
due to magnetization motion is
\begin{equation}
\dot{E}=-2\mathcal R(\dot{\mathbf{m}}_1,\dot{\mathbf{m}}_2,\cdots,
\dot{\mathbf{m}}_N)=-\sum_{i,j}R_{ij}\dot{\mathbf m}_i\cdot\dot{\mathbf m}_j.\label{Erate}
\end{equation}
Before proceeding, we discuss the mathematical properties of the
dissipation matrix $\mathbf R$ and corresponding physical meanings.
Firstly, for a particular motion of $\dot{\mathbf m}_i$, the energy
dissipation rate, which is a physically observable quantity, must be
unique, indicating the uniqueness of every matrix element $R_{ij}$.
Secondly, according to the second law of thermodynamics, the energy
of a system without any energy source must always decrease.
In another word, $\dot E$ is always negative for an arbitrary motion
$\dot{\mathbf m}_i$ indicating that all the elements of the
dissipation matrix $\mathbf R$ must be real and positive.
Thirdly, if sublattices are all equivalent with one another
in an AFM, one has the identical diagonal matrix element,
$R_{ii}=R_{jj}$ for arbitrary $i$ and $j$. Lastly, the permutation
symmetry of AFM sublattices and the action-reaction law both
require the dissipation matrix being symmetric, $R_{ij}=R_{ji}$.
Furthermore, the real symmetric matrix is also consistent with
the requirement of real eigenvalues of $R$.

Since the dissipation matrix $\mathbf R$ is real and symmetric, one
can always find an orthogonal matrix $\mathbf U$ to diagonalize
$\mathbf R$, i.e. $\mathbf U^{\dagger}\mathbf R\mathbf U=diag(\alpha_1,\alpha_2,\cdots,\alpha_N)$.
Thus Eq.~\eqref{Erate} can be rewritten as
\begin{equation}
\dot{E}=-\sum_{i=1}^N\alpha_i \dot{\mathbf n}_i^2,\label{Eratediag}
\end{equation}
where $\mathbf n_i=\sum_{j=1}^NU_{ji}\mathbf m_j$ is the linear
combination of $\mathbf m_i$. Since all the diagonal elements
$\alpha_i$ must be real and positive, the dissipation matrix $\mathbf
R$ is positive-definite, and $\mathbf{n}_i$ ($i=1,2,\cdots,N$) are the
natural order parameters of an $N$-sublattice AFM \cite{note02} (See Appendix
C for details).

Let us compare our result Eq.~\eqref{nsub} with the present theories in
literature. For a FM with its magnetization as the only order parameter,
Eq.~\eqref{nsub} with $N=1$ recovers the LLG equation for a FM.
For $N=2$, $\mathbf R$ is a $2\times 2$ matrix defined by two real positive
numbers, $\alpha$ and $\alpha_{c}$ for the diagonal and off-diagonal
elements,
\begin{eqnarray}
\mathbf R=\left(\begin{array}{cc} \alpha & \alpha_c \\ \alpha_c & \alpha \end{array}\right).\label{R2}
\end{eqnarray}
The matrix defines two order parameters $(\mathbf{m}_1+\mathbf{m}_2)
/\sqrt{2}$ and $(\mathbf{m}_1-\mathbf{m}_2)/\sqrt{2}$, which are the
well-known net magnetization and the N{\'e}el order parameter.
Two corresponding eigenvalues $\alpha \pm \alpha_c$ are the damping
coefficients associated with the motion of the two order parameters.
The Kittel's AFM theory \cite{Kittel1951,Keffer1952} and its extension
to include the dissipative torque corresponds to $\alpha\neq 0$ and
$\alpha_c=0$. The Bar'yakhtar approach is the special case of
$\alpha=\alpha_c$ \cite{Baryakhtar201301,Baryakhtar201302}, which is
not true in general (See Appendix D for details). The {\it ad hoc} damping terms
added into the dynamic equations by Hals {\it et al.}~\cite{Hals2011}
are justified by our results with the correspondence $\alpha_m=(\alpha
+\alpha_c)/2$ and $\alpha_n =(\alpha-\alpha_c)/2$, respectively.
Therefore, our result, Eq.~\eqref{nsub} essentially unifies the
existing phenomenological theories in literature.

The new dissipative torques in Eq.~\eqref{nsub}, $R_{ij}\mathbf m_i
\times\dot{\mathbf{m}}_j$ ($i\neq j$), can also be viewed as an
effective torque on spin $i$ dragged by the motion of spin $j$.
This is similar to the motion of a particle in a fluid where the
motion of neighboring particles can exert a force on the particle.
One can also interpret the torque as the inter-sublattice spin pumping
effect \cite{ly2017}: the motion of $\mathbf{m}_j$ pumps a spin current
of $\alpha_{\rm sp} \mathbf{m}_j\times \dot{\mathbf{m}}_j$, which is
absorbed by $\mathbf m_i$ and results in an effective damping torque
on $\mathbf{m}_i$ of the form $\alpha_c\mathbf{m}_i\times [\mathbf
{m}_i\times (\mathbf{m}_j \times \dot{\mathbf{m}}_j)]\approx \alpha_c
\mathbf{m}_i\times \dot{\mathbf{m}}_j$. $\alpha_c$ measures the
magnitude of the spin pumping. In addition, the spin pumping from
$\mathbf{m}_i$ can enhance its own damping to be $(\alpha_0+
\alpha_c)\mathbf{m}_i \times\dot{\mathbf{m}}_i$, where $\alpha_0$
is the intrinsic Gilbert damping. This consideration leads to
$R_{ii}=\alpha_0 + \alpha_c$ and $R_{ij}= \alpha_c$, and
$R_{ii} > R_{ij} > 0\ (i\neq j)$. For Mn-based metallic AFMs,
recent first-principles calculations show that the magnitude of
the diagonal and off-diagonal dissipation matrix elements are very
close to each other. It implies that the inter-sublattice spin
pumping is the dominant mechanism of damping in bulk metallic AFMs.
It is also interesting to note that the new dissipative torques
$R_{ij}\mathbf m_i\times\dot{\mathbf m}_j$ ($i\neq j$) plays an
important role in the interfacial spin pumping \cite{Akash2017}.

\section{Influence of new torque on magnon lifetime and domain wall velocity}
\begin{figure}
\centering
\includegraphics[width=\columnwidth]{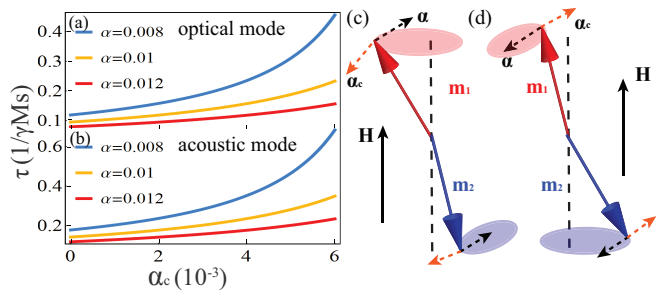}
\caption{(color online) Magnon lifetime as a function of the new
dissipative torque strength $\alpha_c$ for the optical mode (a) and
acoustic mode (b), respectively. The corresponding dynamic modes
are illustrated in (c) and (d). The dashed arrows indicate the
directions of $\alpha$-torque (black) and $\alpha_c$-torque (orange).
Parameters $H_{E} = 858 M_s,\ H_{\rm an}= 14M_s,\ H_0=31M_s,\vert
\mathbf m_1^0\vert=\vert \mathbf m_2^0 \vert$ are used.}
\label{fig1}
\end{figure}

We consider two examples, magnon
lifetime and DW velocity, to highlight the importance of the new
dissipative $\alpha_c$-torque. According to Eq.~\eqref{nsub} and
Eq.~\eqref{R2}, the dynamic equations of a two-sublattice AFM are
\begin{equation}
\begin{aligned}
\dot{\mathbf{m}}_1=-\mathbf{m}_1\times\mathbf{h}_1+\mathbf{m}_1 \times
\left(\alpha\dot{\mathbf{m}}_1+\alpha_c\dot{\mathbf{m}}_2\right ),\\
\dot{\mathbf{m}}_2=-\mathbf{m}_2\times\mathbf{h}_2+\mathbf{m}_2 \times
\left(\alpha\dot{\mathbf{m}}_2+ \alpha_c\dot{\mathbf{m}}_1\right ).
\label{2sub}
\end{aligned}
\end{equation}
In the following, we will show that the $\alpha_c$-torque can
significantly increase the magnon lifetime in an AFM.
We consider a spin wave of the wavevector $\mathbf{k}$ and frequency
$\omega$ as $\mathbf{m}_1=\mathbf{m}_1^0+\delta \mathbf{m}_1
\exp (i\mathbf{k \cdot r} - i \omega t),\ \mathbf{m}_2=\mathbf{m}_2^0
+ \delta \mathbf{m}_2 \exp (i\mathbf{k \cdot r} - i \omega t)$,
where $\mathbf{m}_i^0$ are the magnetic moment of sublattice $i$
in the ground state and $\delta\mathbf m_i$ is a small deviation
perpendicular to $\mathbf m_i$. Following the standard Kittel
approach \cite{Kittel1951,Keffer1952}, we can determine the spin
wave dispersion using the linearized equations of Eq.~(\ref{2sub})
for $\delta\mathbf{m}_1$ and $\delta \mathbf{m}_2$.
The magnon frequency is obtained by solving the secular equation
\begin{equation}
\omega^2 \pm (a_{11} + a_{22}) \omega + a_{11}a_{22} + a_{12}a_{21}=0,\label{quad}
\end{equation}
where $a_{11}=H_0+H_{\rm an} m_1^0+H_{E}m_2^0 -i\omega\alpha m_1^0,\
a_{22}=H_0-H_{\rm an} m_2^0 - H_{E} m_1^0 + i\omega \alpha m_2^0,\
a_{12}=(H_E -i\omega \alpha_c)m_1^0,\ a_{21}=a_{12}m_2^0/m_1^0$.
Here we explicitly write the effective fields as the sum of the
external field $H_0$ along the easy axis, the exchange field
$H_E$ and the anisotropy field $H_{\rm an}$. The two solutions of
Eq.~\eqref{quad} correspond to the acoustic and optical modes of
magnon excitation, respectively. The magnon lifetime
$\tau=-1/\mathrm{Im} (\omega)$ is plotted in Fig.~\ref{fig1} as a
function of $\alpha_c$. The lifetimes of both optical and acoustic
magnons increase dramatically with $\alpha_c$ and the enhancement
is particularly large for small $\alpha$.
This is because the new $\alpha_c$-torque, whose effect is
opposite to the conventional damping torque (the $\alpha$-term),
drags the magnetization away from its equilibrium state
as schematically illustrated in Fig.~\ref{fig1}(c) and (d).

To see the influence of $\alpha_c$ on DW motion, we rewrite
Eq.~(\ref{2sub}) in terms of net magnetization $\mathbf{m}=\mathbf{m}_1
+ \mathbf{m}_2$ and N\'{e}el order
parameter $\mathbf{n}  = \mathbf{m}_1 - \mathbf{m}_2$ that are the natural
order parameters as we discussed after Eq.~\eqref{R2}.
The equations are
\begin{equation}
\begin{aligned}
&\dot{\mathbf{m}}= -\mathbf{m} \times \left(\mathbf{h}_m-\alpha_m \dot{\mathbf{m}}\right) -\mathbf{n} \times \left(\mathbf{h}_n - \alpha_n \dot{\mathbf{n}}\right),\\
&\dot{\mathbf{n}}= -\mathbf{m} \times \left(\mathbf{h}_n-\alpha_n \dot{\mathbf{n}}\right)-\mathbf{n} \times\left( \mathbf{h}_m -\alpha_m\dot{\mathbf{m}}\right),
\label{llgmn}
\end{aligned}
\end{equation}
where $\alpha_m=(\alpha+\alpha_c)/2$, $\alpha_n=(\alpha-\alpha_c)/2$,
$\mathbf{h}_m =-\delta \mathcal{H}/\delta \mathbf{m}$,
$\mathbf{h}_n =-\delta \mathcal{H}/\delta \mathbf{n}$ are
respectively the effective magnetization field and the N\'{e}el
field and $\mathcal H$ is the free energy of the system.
For small deviation $\mathbf{m}_1 \approx - \mathbf{m}_2$, $|\mathbf{m}|
\ll |\mathbf{n}|$ and the magnitude of $\mathbf{n}$ is nearly conserved.
We keep only the terms that preserve $|\mathbf{n}|$ and that are linear
in $\mathbf{m}$, the dynamic equations can be further simplified as
\begin{equation}
\begin{aligned}
&\dot{\mathbf{m}}= -\mathbf{n} \times \left(\mathbf{h}_n - \alpha_n \dot{\mathbf{n}}\right),\\
&\dot{\mathbf{n}}= -\mathbf{n} \times\left( \mathbf{h}_m -\alpha_m\dot{\mathbf{m}}\right).
\label{sim}
\end{aligned}
\end{equation}
Note that $\mathbf h_m$ is of the same order of $\mathbf m$ \cite{Tveten2013}.
It is worth mentioning that Eq.~\eqref{sim} is the same as the equation
used by Hals {\it et al.} \cite{Hals2011} with similar assumptions.

For a uniaxial 1D AFM, the free energy density functional is
$\mathcal{H}= H_{E} \mathbf{m}^2/2 + A(\partial_z\mathbf{n})^2/2
-H_{\rm an} n_z^2/2 -  \mathbf{b \cdot n}$,
where $A$ is the inhomogeneous exchange constant and $\mathbf{b}$
is the N\'{e}el field generated by an electric current through
spin-orbit interaction \cite{Zelezny2014, Helen2016}.
Eliminating $\mathbf{m}$ from Eq. (\ref{sim}), the decoupled
dynamic equation of $\mathbf{n}$ is \cite{Hals2011}
\begin{equation}
\mathbf{n} \times \left (  -\ddot {\mathbf{n}}+ \alpha_m \dot{\mathbf{h}}_n + H_{E} \mathbf{h}_n- \alpha_n H_{E} \dot{\mathbf{n}}\right ) =0.
\end{equation}

For the steady motion of a rigid DW of form $\mathbf{n}(z)=\mathbf{n}
(z-vt)$, DW velocity can be analytically obtained (See Appendix E for details),
\begin{equation}
v = \frac{ bc\Delta_0}{\sqrt{(\alpha_{\mathrm{eff}})^2 c^2 +
( b \Delta_0)^2}},
\end{equation}
where $\Delta_0 = \sqrt{A/H_{\rm an}}$ is the static DW width,
$c= \sqrt{AH_E}$ is the magnon velocity, and the effective damping
is $\alpha_{\mathrm{eff}} =\alpha_n+\alpha_m H_{\rm an}/(3H_E)$.
Recent first-principles calculations show that $\alpha_m$ is one to
three orders of magnitude larger than $\alpha_n$ for Mn-based metallic
AFMs \cite{ly2017}. Thus, $\alpha_c$-term slows down DW propagation
and enhances greatly the effective damping $\alpha_{\rm eff}$.
DW velocity reduction is particularly strong when the ratio of $H_{\rm
an}$ and $H_E$ is large. Thus to increase the DW velocity, the AFMs
with strong exchange interaction and weak anisotropy are preferred.
We have also performed first-principle calculations for metallic AFM
Mn$_2$Au with the tetragonal structure \cite{Wu2012,Shick2010}
and find $\alpha_m=0.42$ and $\alpha_n=2.8\times 10^{-3}$.
The effective damping is significantly enhanced if $H_{\rm an}
\ge10^{-3}H_E$.

\section{Conclusions}
In conclusion, a proper set of dynamic equations
for AFMs is derived from the Lagrange equation with the Rayleigh
dissipation functional. Our phenomenological theory unifies all the
existing AFM dynamic equations in literature and propose
a general way of constructing the order parameters of magnetic
systems with multiple sublattices. We discover a new anti-damping-like
torque that significantly influences the magnon lifetime and DW velocity.
The new torque naturally explains the recent first-principle results that
the damping coefficient associated with the motion of magnetization is
much larger than that associated with the motion of N\'{e}el order.

\begin{acknowledgments}
H.Y.Y. thanks Jiang Xiao for helpful discussions.
This work was financially supported by National Key Research
and Development Program of China (2017YFA0303300) and National
Natural Science Foundation of China (Grants No. 61704071,
No. 61774017, No. 61774018, No. 11734004 and No. 21421003).
Z.Y. acknowledges the Recruit Program of Global Youth Experts.
X.R.W. was supported by the National Natural Science
Foundation of China (Grant No. 11774296) as well as
Hong Kong RGC Grants No. 16300117 and No. 16301816.
\end{acknowledgments}

\section*{Appendix A: Derivation of the kinetic energy expression}
The kinetic energy of a spin in state $|\mathbf{S}(t) \rangle$ at
time $t$ is
$\mathcal{T} \equiv i\langle \mathbf{S}(t)| \partial_t |\mathbf{S}(t) \rangle$.
For $S= 1/2$ spin, the general expression of $|\mathbf{S} (t)\rangle$ is
\begin{widetext}
\begin{equation}
\begin{aligned}
&|\mathbf{S}(t) \rangle  =\cos \frac{\theta}{2} e^{-i\varphi/2}|\uparrow \rangle
+ \sin \frac{\theta}{2} e^{i\varphi/2}|\downarrow \rangle \\
&\Rightarrow \\
&\partial_t |\mathbf{S}(t) \rangle =\left ( -\frac{1}{2}\partial_t \theta \sin \frac{\theta}{2}
- \frac{i}{2} \partial_t \varphi \cos \frac{\theta}{2} \right )e^{-i\varphi/2}  |\uparrow \rangle
+ \left (\frac{1}{2}\partial_t \theta \cos \frac{\theta}{2} +
\frac{i}{2} \partial_t \varphi \sin \frac{\theta}{2} \right )e^{i\varphi/2}|\downarrow \rangle \\
&\Rightarrow \\
&\langle \mathbf{S}(t)|\partial_t |\mathbf{S}(t) \rangle=-\frac{i}{2} \partial_t \varphi \cos \theta.
\end{aligned}
\end{equation}
\end{widetext}

Therefore $\mathcal{T}$ is
\begin{equation}
\begin{aligned}
\mathcal{T} \equiv i\langle \mathbf{S}(t)| \partial_t |\mathbf{S}(t) \rangle
=\frac{1}{2} \partial_t \varphi \cos \theta.
\end{aligned}
\end{equation}
For general $S$, $\mathcal{T}$ is generalized to
$\mathcal{T} =S \partial_t \varphi \cos \theta$ by considering a spin
coherent state $|\mathbf{S} (t)\rangle$ \cite{Tatara2008}.
The classical counterpart of the spin orientation is
$\mathbf{S}=\sqrt{S(S+1)}\mathbf{m}\approx S\mathbf{m}$, where
$\mathbf{m}$ is a unit vector. The kinetic energy could be rewritten as
$\mathcal{T} = S \cot \theta \mathbf{\hat{\varphi}}\cdot \partial_t \mathbf{m}
=-S \mathbf{A}\cdot \partial_t \mathbf{m}$,
where $\nabla_\mathbf{m} \times \mathbf{A} = \nabla_\mathbf{m} \times (\cot \theta \hat{\varphi})
=  \mathbf{m}$. In the dimensionless form, one has
$\mathcal{T}=\mathbf A(\mathbf{m}) \cdot \partial_t \mathbf{m}$.

For a magnetic lattice with spin $\mathbf{S}_i$ on $i-$th lattice site,
the total kinetic energy is $\mathcal{T}_t=\sum_i \mathcal{T}_i=
\sum_i \mathbf{A}_i(\mathbf{m}_i) \cdot \partial_t \mathbf{m}_i$
In the continuous limit, $\mathcal{T}_t = 1/d^3\int \mathbf{A}(\mathbf{m})
\cdot \partial_t \mathbf{m}dV$ where $d$ is lattice constant.
The kinetic energy density is $\mathcal{T}=\mathbf{A}(\mathbf{m})
\cdot \partial_t \mathbf{m}$.

\section*{Appendix B: Derivation of the dynamic equation}
We start from the Lagrange equation,
\begin{equation}
-\frac{\delta \mathcal{T}}{\delta \mathbf{m}}
+ \frac{\partial }{\partial t}\frac{\delta \mathcal{T}}{\delta \dot{\mathbf{m}}}
+\left ( -\mathbf{h} + \frac{\delta \mathcal{R}}{\delta \dot{\mathbf{m}}} \right ) =0,
\label{lagg-2}
\end{equation}
where $\mathcal{R}$ is the Rayleigh functional of the magnetic system,
$\mathbf{h}$ is the effective field acting on the magnetic moment $\mathbf{m}$,
$\dot{\mathbf{m}} \equiv \partial_t \mathbf{m}$.
$\mathbf{m} \times$ Eq. (\ref{lagg-2}) gives
\begin{equation}
\mathbf{m} \times \left (\frac{\partial }{\partial t}\frac{\delta \mathcal{T}}{\delta \dot{\mathbf{m}}}
-\frac{\delta \mathcal{T}}{\delta \mathbf{m}} \right )-\mathbf{m} \times
\left ( \mathbf{h} - \frac{\delta \mathcal{R}}{\delta \dot{\mathbf{m}}} \right ) =0.
\label{lag-2}
\end{equation}
Using the relations
\begin{equation}
\mathcal{T}=\mathbf{A}(\mathbf{m}) \cdot \frac{\partial \mathbf{m}}{\partial t},\, \, \, \mathrm{and}\,\,\,
\nabla_\mathbf{m} \times \mathbf{A} =\mathbf{m},
\end{equation}
we have
\begin{equation}
-\frac{\delta \mathcal{T}}{\delta m_i}=-\frac{\partial \mathbf{A}}{\partial m_i}
\cdot \frac{\partial \mathbf{m}}{\partial t}, \frac{\partial}{\partial t}
\frac{\delta \mathcal{T}}{\delta \dot{m}_i}=\frac{\partial A_i}{\partial t},\,\mathrm{for}\,i=1,2,3.
\end{equation}
Then we have
\begin{equation}
\begin{aligned}
\left [ \mathbf{m} \times \left (\frac{\partial }{\partial t}
\frac{\delta \mathcal{T}}{\delta \dot{\mathbf{m}}} -
\frac{\delta \mathcal{T}}{\delta \mathbf{m}}\right )\right ]_i
&=\epsilon_{ijk}m_j \left ( \frac{\partial A_k}{\partial t}
- \frac{\partial A_p}{\partial m_k} \frac{m_p}{\partial t} \right ) \\
&=\epsilon_{ijk}m_j \left ( \frac{\partial A_k}{\partial m_p} \frac{\partial m_p}{\partial t}
- \frac{\partial A_p}{\partial m_k} \frac{m_p}{\partial t} \right ) \\
&=\epsilon_{ijk}m_j \left ( \frac{\partial A_k}{\partial m_p}
- \frac{\partial A_p}{\partial m_k} \right ) \frac{\partial m_p}{\partial t}.\\
\end{aligned}
\end{equation}
For $i=1$,
\begin{widetext}
\begin{equation}
\begin{aligned}
\left [ \mathbf{m} \times \left (\frac{\partial }{\partial t}
\frac{\delta \mathcal{T}}{\delta \dot{\mathbf{m}}} -\frac{\delta \mathcal{T}}{\delta \mathbf{m}}\right )\right ]_1
&=\epsilon_{1jk}m_j \left ( \frac{\partial A_k}{\partial m_p}
- \frac{\partial A_p}{\partial m_k} \right ) \frac{\partial m_p}{\partial t}\\
&=\epsilon_{123}m_2 \left ( \frac{\partial A_3}{\partial m_p}
- \frac{\partial A_p}{\partial m_3} \right ) \frac{\partial m_p}{\partial t} +
\epsilon_{132}m_3 \left ( \frac{\partial A_2}{\partial m_p}
- \frac{\partial A_p}{\partial m_2} \right ) \frac{\partial m_p}{\partial t}\\
&=m_2 \left ( \frac{\partial A_3}{\partial m_1}  - \frac{\partial A_1}{\partial m_3} \right )
\frac{\partial m_1}{\partial t} +m_2 \left ( \frac{\partial A_3}{\partial m_2}
- \frac{\partial A_2}{\partial m_3} \right ) \frac{\partial m_2}{\partial t} \\
&-m_3 \left ( \frac{\partial A_2}{\partial m_1}  - \frac{\partial A_1}{\partial m_2} \right )
\frac{\partial m_1}{\partial t}-m_3 \left ( \frac{\partial A_2}{\partial m_3}
- \frac{\partial A_3}{\partial m_2} \right ) \frac{\partial m_3}{\partial t},
\underline{use\ \nabla_\mathbf{m} \times \mathbf{A} =\mathbf{m}}\\
&=-m_2^2\frac{\partial m_1}{\partial t} +m_2m_1\frac{\partial m_2}{\partial t}
-m_3^2\frac{\partial m_1}{\partial t} +m_3m_1\frac{\partial m_3}{\partial t}\\
&=(m_1^2 - 1)\frac{\partial m_1}{\partial t} + \frac{1}{2}m_1 \frac{\partial m_2^2}{\partial t}
+ \frac{1}{2}m_1 \frac{\partial m_3^2}{\partial t}\\
&=(m_1^2 - 1)\frac{\partial m_1}{\partial t} + \frac{1}{2}m_1\frac{\partial (1-m_1^2)}{\partial t} \\
&=-\frac{\partial m_1}{\partial t}.
\end{aligned}
\end{equation}
\end{widetext}

Similarly, we have
\begin{equation}
\begin{aligned}
\left [ \mathbf{m} \times \left (\frac{\partial }{\partial t}
\frac{\delta \mathcal{T}}{\delta \dot{\mathbf{m}}}
 -\frac{\delta \mathcal{T}}{\delta \mathbf{m}}\right )\right ]_i
&=\frac{\partial m_i}{\partial t}, i=2,3.
\end{aligned}
\end{equation}

Finally, we arrive at the dynamic equation
\begin{equation}
\frac{\partial \mathbf{m}}{\partial t} =-\mathbf{m} \times
\left ( \mathbf{h} - \frac{\delta \mathcal{R}}{\delta \dot{\mathbf{m}}} \right ).
\end{equation}

\section*{Appendix C: Order parameters and damping for $N$=3 and 4}
In this section, we show the well-defined order parameters of a $N$-sublattice AFM ($N=3,4$) from the diagonalization of dissipation matrix. For $N=3$, the dissipation matrix reads
\begin{equation}
\mathbf{R}=
\left ( \begin{array}{ccc}
\alpha & \alpha_c & \alpha_c\\
\alpha_c  & \alpha & \alpha_c\\
\alpha_c  & \alpha_c & \alpha\\
\end{array} \right ).
\end{equation}
The eigenvalues and the corresponding eigenvectors
are
\begin{equation}
\begin{aligned}
&\alpha_1 = \alpha-\alpha_c,v_1 =\frac{1}{\sqrt{2}} (-1,1,0),\\
&\alpha_2 = \alpha-\alpha_c,v_2 = \frac{1}{\sqrt{2}}(-1,0,1),\\
&\alpha_3 = \alpha+2\alpha_c,v_3 = \frac{1}{\sqrt{3}}(1,1,1).\\
\end{aligned}
\end{equation}
The dissipation matrix $\mathbf{R}$ can be diagonalized as $\mathbf{\Lambda} = diag (\alpha_1,\alpha_2,\alpha_3)$ under the
orthogonal transformation $\mathbf{U}^T\mathbf{R}\mathbf{U}$ with $\mathbf{U} = (v_1^T,v_2^T,v_3^T)$.
Then the order parameters of a $3$-sublattice AFM should be
\begin{equation}
\begin{aligned}
&\mathbf{n}_1 = \frac{1}{\sqrt{2}}(-\mathbf{m}_1 + \mathbf{m}_2),\\
&\mathbf{n}_2 = \frac{1}{\sqrt{2}}(-\mathbf{m}_1 + \mathbf{m}_3),\\
&\mathbf{n}_3 = \frac{1}{\sqrt{3}}(\mathbf{m}_1 + \mathbf{m}_2 + \mathbf{m}_3).
\end{aligned}
\end{equation}

For $N=4$, the dissipation matrix reads
\begin{equation}
\mathbf{R}=
\left ( \begin{array}{cccc}
\alpha & \alpha_c & \alpha_c& \alpha_c\\
\alpha_c  & \alpha & \alpha_c& \alpha_c\\
\alpha_c  & \alpha_c & \alpha& \alpha_c\\
\alpha_c  & \alpha_c & \alpha_c & \alpha
\end{array} \right ),
\end{equation}
with the eigenvalues and eigenvectors,
\begin{equation}
\begin{aligned}
&\alpha_1 = \alpha-\alpha_c,v_1 =\frac{1}{2} (-1,1,-1,1),\\
&\alpha_2 = \alpha-\alpha_c,v_2 =\frac{1}{2} (-1,-1,1,1),\\
&\alpha_3 = \alpha-\alpha_c,v_3 =\frac{1}{2} (-1,1,1,-1),\\
&\alpha_4 = \alpha+3\alpha_c,v_4 = \frac{1}{2}(1,1,1,1).
\end{aligned}
\end{equation}
Similarly, the order parameters of a 4-sublattice AFM should be
\begin{equation}
\begin{aligned}
&\mathbf{n}_1 = \frac{1}{2}(-\mathbf{m}_1 + \mathbf{m}_2  -\mathbf{m}_3 + \mathbf{m}_4),\\
&\mathbf{n}_2 = \frac{1}{2}(-\mathbf{m}_1 - \mathbf{m}_2 + \mathbf{m}_3 + \mathbf{m}_4),\\
&\mathbf{n}_3 = \frac{1}{2}(-\mathbf{m}_1 + \mathbf{m}_2 + \mathbf{m}_3 - \mathbf{m}_4),\\
&\mathbf{n}_4 = \frac{1}{2}(\mathbf{m}_1 + \mathbf{m}_2 + \mathbf{m}_3 + \mathbf{m}_4).
\end{aligned}
\end{equation}
Our theory provides a systematic justification for the order parameters conjectured in the literature \cite{Gomonay2012}.

\section*{Appendix D: Incompleteness of the Bar'yakhtar approach}
In this section, we use a two-sublattice AFM show the incompleteness of
Bar'yakhtar's approach \cite{Baryakhtar1984,Baryakhtar1988,
Baryakhtar201301,Baryakhtar201302} for the dissipative torques.
The dynamic equation of the two-sublattice AFM is,
\begin{eqnarray}
\dot{\mathbf{m}}_i= -\mathbf{m}_i \times \mathbf{h}_i + \mathbf{R}_i,
\label{bar1}
\end{eqnarray}
where $\mathbf{R}_i = \delta q /\delta \mathbf{h}_i$ is the
corresponding dissipative torque on the $i$-th sublattice and $q$ is the
Bar'yakhtar dissipation function \cite{Baryakhtar1984, Baryakhtar1988,
Baryakhtar201301, Baryakhtar201302}. The original form of $q$ is
constructed based on the symmetry of the lattice and the conservation
of total magnetization $\mathbf{m}=\mathbf{m}_1+\mathbf{m}_2$.
Following Refs.~\cite{Baryakhtar201301} and \cite{Baryakhtar201302},
we choose $q$ arising from both exchange interaction and anisotropy
without loss of generality
\begin{eqnarray}
q= \frac{1}{2} \left[\Lambda_-(\mathbf{h}_1 - \mathbf{h}_2)^2+\Lambda_z\left(h^2_{1,z}+h^2_{2,z}\right)\right].
\label{bar2}
\end{eqnarray}
By substituting Eq.~(\ref{bar2}) into Eq.~(\ref{bar1}), the dynamic equation becomes
\begin{equation}
\begin{aligned}
\dot{\mathbf{m}}_1= -\mathbf{m}_1 \times \mathbf{h}_1+ \Lambda_- \mathbf{h}_1  - \Lambda_-  \mathbf{h}_2+\Lambda_z h_{1,z}\hat z,\\
\dot{\mathbf{m}}_2= -\mathbf{m}_2 \times \mathbf{h}_2 - \Lambda_- \mathbf{h}_1  + \Lambda_-  \mathbf{h}_2+\Lambda_z h_{2,z}\hat z.
\label{b2sub}
\end{aligned}
\end{equation}
For small deviation from the equilibrium state, i.e. $\mathbf m_i$
along the $z$-axis, we have
\begin{equation}
\mathbf m_1=\hat z+\delta \mathbf m_1e^{i\mathbf k\cdot\mathbf r-i\omega t},~\mathbf m_2=-\hat z+\delta \mathbf m_2e^{i\mathbf k\cdot\mathbf r-i\omega t},
\end{equation}
where $\delta\mathbf m_i$ is the transverse component of $\mathbf m_i$. We immediately see the $\Lambda_z$ terms corresponding to the
longitudinal relaxation of the magnetization. However, the magnitude of
the magnetic moment can hardly vary in magnetization dynamics except
for some extreme cases like the laser-induced demagnetization \cite{Atxitia2017}.
Neglecting the longitudinal relaxation, we use the approximation made by Bar'yakhtar {\it et al.}\cite{Baryakhtar1984}, i.e. $\mathbf{h}_i = \mathbf{m}_i \times \dot{\mathbf{m}}_i$ and $\mathbf{m}_1 \approx -\mathbf{m}_2$. Then Eq.~(\ref{b2sub}) is reduced to
\begin{equation}
\begin{aligned}
\dot{\mathbf{m}}_1= -\mathbf{m}_1 \times \mathbf{h}_1+ \mathbf{m}_1 \times \left ( \Lambda_- \dot{\mathbf{m}}_1+\Lambda_- \dot{\mathbf{m}}_2 \right ),\\
\dot{\mathbf{m}}_2= -\mathbf{m}_2 \times \mathbf{h}_2+ \mathbf{m}_2 \times \left ( \Lambda_- \dot{\mathbf{m}}_2+\Lambda_- \dot{\mathbf{m}}_1 \right ).
\label{2sub-1}
\end{aligned}
\end{equation}
Through linear combination of the two equations, the dynamic equations of $\mathbf{m}$ and $\mathbf{n}$ up to the linear orders of $\mathbf{m}$ are derived as,
\begin{equation}
\begin{aligned}
&\dot{\mathbf{m}}= -\mathbf{n} \times  \mathbf{h}_n , \\
&\dot{\mathbf{n}}=-\mathbf{n} \times  \mathbf{h}_m  + \alpha_m \mathbf{n} \times \dot{\mathbf{m}},
\end{aligned}
\label{llgmn}
\end{equation}
where $\alpha_m=\Lambda_-$. This set of equations does not include the the damping associated with the motion of N\'{e}el order or equivalently $\alpha_n=0$ in contrast to the literature \cite{Hals2011}. In addition, recent first-principles calculation has demonstrated that $\alpha_n$ is finite for Mn-based AFMs.

If we release the constraint of magnetization conservation, we can generalize the Bar'yakhtar dissipation function to be
\begin{equation}
q= \frac{\Lambda_-}{2} (\mathbf{h}_1 - \mathbf{h}_2)^2 + \frac{\Lambda_+}{2} (\mathbf{h}_1+ \mathbf{h}_2)^2,
\end{equation}
Then the dynamic equations become
\begin{widetext}
\begin{equation}
\begin{aligned}
\dot{\mathbf{m}}_1= -\mathbf{m}_1 \times \mathbf{h}_1+ \left ( \Lambda_- +\Lambda_+\right)\mathbf{m}_1 \times \dot{\mathbf{m}}_1+\left(\Lambda_- - \Lambda_+\right)\mathbf{m}_1 \times \dot{\mathbf{m}}_2 ,\\
\dot{\mathbf{m}}_2= -\mathbf{m}_2 \times \mathbf{h}_2+\left(\Lambda_- -\Lambda_+\right)\mathbf{m}_2 \times \dot{\mathbf{m}}_1 + \left ( \Lambda_- +\Lambda_+\right)\mathbf{m}_2 \times \dot{\mathbf{m}}_2.
\label{2sub-2}
\end{aligned}
\end{equation}
\end{widetext}

The above equations could reproduce our key results in the main text with $\alpha=\Lambda_-+\Lambda_+$ and $\alpha_c=\Lambda_--\Lambda_+$, or equivalently, $\alpha_m=\Lambda_-$ and $\alpha_n=\Lambda_+$.
\section*{Appendix E: Derivation of Domain Wall velocity}
We consider a uniaxial one-dimensional system with the energy density functional
$\mathcal{H}= H_{E} \mathbf{m}^2/2 + A(\partial_z\mathbf{n})^2/2
-H_{\rm an} n_z^2/2 -  \mathbf{b} \cdot {\mathbf n},$ where $H_E$ and $H_{\rm an}$
are exchange field and anisotropy field, respectively,
$A$ is the inhomogeneous exchange constant, $\mathbf{b}$
is the N\'{e}el field generated by an electric current through
spin-orbit interaction \cite{Zelezny2014, Helen2016}, $z$-axis
is the length direction of the nanowire. The dynamics of N\'{e}el
order $\mathbf{n}$ is determined by the equation
\begin{equation}
\mathbf{n} \times \left (  -\ddot{\mathbf{n}}
+ \alpha_m \dot{\mathbf{h}}_n + H_{E} \mathbf{h}_n
- \alpha_n H_{E} \dot{\mathbf{n}}\right ) =0,
\label{ne}
\end{equation}
where $\mathbf{h}_n=-\delta \mathcal{H}/\delta \mathbf{n}$ is the
effective field on the N\'{e}el order $\mathbf{n}$, $\alpha_m$ and $\alpha_n$
are respectively the damping coefficients of magnetization $\mathbf{m}$
and N\'{e}el order $\mathbf{n}$. In spherical coordinates,
$\mathbf{n}=(\sin \theta \cos \varphi, \sin \theta \sin \varphi, \cos \theta)$,
where $\theta$ and $\varphi$ are the polar and azimuthal angles of the
N\'{e}el order, respectively, then the dynamic Eq.~(\ref{ne}) can be reduced to
\begin{widetext}
\begin{eqnarray}
&&-\partial_{tt} \theta - H_E\alpha_n \partial_t \theta + \sin \theta \cos \theta (\partial_{t} \theta)^2
+ H_E[-b\sin \theta -H_{\rm an} \sin \theta \cos \theta +A \partial_{zz} \theta ]+ \nonumber \\
&&\alpha_m [H_{\rm an} \partial_t \theta \sin^2\theta + A \partial_t \partial_{zz} \theta
- A (\partial_z \theta)^2 \partial_t \theta]=0, \label{theta1} \\
&&(\sin \theta \partial_{tt} \varphi + 2 \cos \theta \partial_t \theta \partial_t \varphi
+ H_E\alpha_n \sin \theta \partial_t \varphi) + \alpha_m (A \sin \theta \partial_t \varphi
(\partial_z \theta)^2 - A \cos \theta \partial_t \varphi \partial_{zz} \theta)=0. \label{theta2}
\end{eqnarray}
\end{widetext}

For the steady motion of DWs, the magnetic structure has translational symmetry, such that,
\begin{equation}
\theta (z, t) =\theta ( z- vt),
~\varphi(z, t)= \varphi(z - vt).
\end{equation}
Hence $\partial_t \theta = -v \partial_z\theta, \partial_t \varphi = -v \partial_z \varphi$.
Similar to ferromagnetic DW counterpart, we consider the case that
the DW plane tilts to a constant angle, such that $\partial_t \varphi= 0$,
then Eq. (\ref{theta2}) is satisfied automatically while
Eq. (\ref{theta1}) is reduced to
\begin{widetext}
\begin{eqnarray}
&&-v^2 \partial_{zz} \theta +v H_E\alpha_2 \partial_z \theta
+ \sin \theta \cos \theta (\partial_{t} \theta)^2
+ H_E[-b\sin \theta -H_{\rm an} \sin \theta \cos \theta
+A \partial_{zz} \theta ] \nonumber \\
&&+\alpha_m [-v H_{\rm an} \partial_z \theta \sin^2\theta
- Av \partial_t \partial_{zz} \theta
+ Av (\partial_z \theta)^3 ]=0. \label{theta3}
\end{eqnarray}
\end{widetext}

By multiplying both sides of Eq. (\ref{theta3}) by $\partial_z \theta$ and
integrating on the sample region, we obtain the DW velocity $v$ as
\begin{equation}
v=\frac{2 b H_E}{\alpha_n H_E \int (\partial_z \theta)^2 dz
+ \alpha_m \int F(\theta) d \theta},
\end{equation}
where $F(\theta) = A(\partial_z \theta)^3
-H_{\rm an} \sin^3\theta \partial_z \theta - A \partial_{zzz} \theta$.
In the small-field regime, the Walker solution is a good approximation, then
the integral can be calculated analytically and the velocity becomes
\begin{equation}
v \approx \frac{ b \Delta}{\alpha_n + \alpha_m A/(3H_E\Delta^2)}.
\label{vfm}
\end{equation}

Different from the ferromagnetic case where DW velocity first increases
with external field and then decreases beyond the Walker breakdown field,
there is no breakdown for antiferromagnetic DW propagation and the DW
velocity keeps increasing with the N{\'e}el field up to the magnon
velocity $c= \sqrt{AH_E}$ \cite{Helen2016}.
Physically, the difference is due to the fact that the DW plane of two
sublattices in an AFM rotates to the opposite direction and results
in the zero tilting angle of N\'{e}el order ($\varphi=0$), such that
the limitation from the hard-axis anisotropy disappears.
Mathematically, the DW width is significantly reduced as the
velocity increases in the way that $\Delta = \Delta_0 \sqrt{1-
v^2/c^2}$ \cite{Shiino2016}, where $\Delta_0 = \sqrt{A/H_{\rm an}}$
is the DW width at zero fields. By substituting $\Delta$ into Eq.
(\ref{vfm}), the velocity can be obtained as
\begin{equation}
v =  \frac{bc\Delta_0}{\sqrt{[\alpha_n + \alpha_m A/(3H_E\Delta_0^2)]^2c^2 + b^2 \Delta_0^2}}.
\end{equation}


\begin{thebibliography}{}
\bibitem{note_nu} In an AFM with noncollinear magnetization, the dissipation
depends also on the magnetization gradient $\nabla \mathbf{m}_i$. 
This effect will modify our current dissipation matrix by a quantity proportional to the magnetization gradient (domain wall widths). 
Nevertheless, such effect does not change the physics presented in our work,
and at most renormalizes the parameters in the dynamical equations such as
the Gilbert damping coefficient.
\bibitem{note02} For the case that the eigegvalues are degenerate,
we need to make linear superposition of the eignvectors to make them orthogonal.
\end{thebibliography}
\end{document}